\documentclass[12pt]{article}

\usepackage{mathrsfs}
\usepackage{amsfonts}
\usepackage{graphicx}
\usepackage[crop=pdfcrop]{pstool}
\usepackage{color}
\usepackage{chet}

\date{June 2011}

\preprint{UCSD-PTH-11-11}

\title{Scale without Conformal Invariance:\\\vspace{6pt} An Example}

\author{Jean-Fran\c{c}ois Fortin, Benjam\'\i{}n
Grinstein and Andreas Stergiou\emails{jffortin@physics.ucsd.edu,
bgrinstein@ucsd.edu, stergiou@physics.ucsd.edu}}

\affiliation{Department of Physics, University of California, San Diego, La
Jolla, CA 92093 USA}

\abstract{We give an explicit example of a model in $D=4-\epsilon$
space-time dimensions that is scale but not conformally invariant, is
unitary, and has finite correlators. The invariance is associated with a
limit cycle renormalization group (RG) trajectory. We also prove, to second
order in the loop expansion, in $D=4-\epsilon$, that scale implies
conformal invariance for models of any number of real scalars. For models
with one real scalar and any number of Weyl spinors we show that scale
implies conformal invariance to all orders in perturbation theory.}

\newcommand{\CQ}{\ensuremath{\mathcal{Q}}}
\newcommand{\CP}{\ensuremath{\mathcal{P}}}
\DeclareMathOperator{\Rea}{Re}
\DeclareMathOperator{\Ima}{Im}

\newcommand{\SQ}{\mathsf{Q}}
\newcommand{\T}[2]{T_{#2}^{\phantom{#2}\!#1}}

\begin{document}

\maketitle

\begin{center}
  \begin{minipage}{0.8\textwidth}
    \begin{center}
      \textbf{Erratum}
    \end{center}
    \vspace{-10pt}
    The original claim of this paper was that the set of couplings given in
    Eq.~\eqref{ScaleInvPoint} define a theory that is scale invariant
    without being conformal in $D=4-\epsilon$ spacetime dimensions. This
    claim is false, as was later realized by the
    authors~\cite{Fortin:2012hn}. The correct interpretation of
    Eq.~\eqref{ScaleInvPoint} is that the theory defined by these couplings
    is fully conformal, although it lives on a limit cycle of the
    traditional dim-reg beta function when the anomalous dimension matrix
    is chosen to be symmetric. This paper remains posted due to this novel
    feature of the solution \eqref{ScaleInvPoint}. For more details the
    reader is referred to~\cite{Fortin:2012hn}.
  \end{minipage}
\end{center}

\newsec{Introduction} Can a theory display a symmetry under
dilatations but not be invariant under conformal transformations?  The
answer to the converse, whether a theory can be invariant under
conformal transformations but not under scaling, has long been
known. The algebra of the conformal group gives the generator of
dilatations in terms of conserved generators of conformal
transformations and translations. Hence, conformal plus Poincar\'{e}
invariance implies dilatation invariance. But whether scale implies
conformal has remained an elusive open question.

Using an argument of Zamolodchikov, Polchinski has proved, in $D=2$
space-time dimensions, that a unitary model cannot be scale-invariant
without also being conformal\rcite{Polchinski:1987dy}. The assumption
of unitarity seems to play an essential role. Indeed, Riva and Cardy
have exhibited a model with $D=2$ Euclidean dimensions with scale but
not conformal symmetry\rcite{Riva:2005gd}. But the model is not
reflection-positive, the Euclidean version of unitarity. An earlier
model by Hull and Townsend\rcite{Hull:1985rc} seems to contradict
Polchinski's result, but this may be attributed to the violation of a
technical assumption in Polchinski's argument, namely the existence of
finite correlators of the stress-energy tensor. More recently other
counterexamples have been given, however every case violates one of
the assumptions of the theorem (unitarity, existence and finiteness of
correlators)\rcite{Iorio:1996ad,Ho:2008nr}.

Polchinski went on to show that, at one-loop order, a scalar field
theory in $D=4-\epsilon$ is necessarily conformal if it is scale
invariant. We will review his argument below and show that it can be
extended to one higher order in the loop expansion. A recent one-loop analysis by Dorigoni and Rychkov \rcite{Dorigoni:2009ra} extended Polchinski's argument to the case of theories with both scalars and fermions. We show that their result can be extended to all orders in perturbation theory for models with an arbitrary number of spin-$\frac12$ fields and no more than one real scalar.

Polchinski also reviewed the literature on the subject to that
date. Let us briefly survey salient, mostly recent work on the
relation between dilatation and conformal invariance since his review.
Antoniadis and Buican studied $\mathcal{N}=1$ supersymmetric theories
with an R-symmetry\rcite{Antoniadis:2011gn}. They proved that any
unitary fixed point is either superconformal or corresponds to a model
that has at least two real non-conserved dimension-two scalar singlet
operators. They also demonstrated that any IR fixed point reached from
a flow from a UV superconformal fixed point is itself superconformal,
provided some technical assumptions are met. Nakayama has taken a
fresh approach to the question. He considers the AdS/CFT
correspondence to claim that the null energy condition in the bulk
gravitational theory guarantees the equivalence in the boundary theory
between scale and conformal
invariance\rcite{Nakayama:2010zz,Nakayama:2010wx,Nakayama:2009fe}.

El-Showk, Nakayama and Rychkov have pointed out that Maxwell theory in $D\ne4$ is scale but not conformally invariant \cite{ElShowk:2011gz}.  Jackiw and Pi have shown that Maxwell's action integral in any $D$ is invariant under conformal transformations if  the field strength tensor $F_{\mu\nu}$ is taken as a primary field of scaling dimension $D/2$, but then Maxwell's equations are not covariant under conformal transformations \cite{Jackiw:2011vz}. We note in passing that if the scaling dimension is taken to be $D-2$, then Maxwell's equations are conformally covariant, but the action integral is not invariant, unless $D=4$.  We hasten to indicate that Maxwell theory is a free field theory, and hence of limited interest. Furthermore, in the presence of sources the model is no longer invariant under dilatations in $D\ne4$.

To summarize the situation until this work, for $D > 2$ there appears to be neither an interacting counterexample nor a proof that scale implies conformal invariance. We have found that, in $D=4-\epsilon$, a model of
two Weyl spinors and two real scalars is invariant under scaling but
not under conformal transformations. More specifically, we will show
that there exist points in the parameter space of the model for which
a combined transformation by scaling and by an internal rotation among
the scalars is a symmetry. We will argue that these points must not be
isolated, but lie on RG trajectories that are scale but not
conformally invariant and, most remarkably, discover that these RG
trajectories form closed loops or display ergodic behavior. To the best of our knowledge neither limit
cycles nor ergodic RG trajectories have ever before been reported for a
relativistic field theory.

In preparation for our analysis, we review the arguments of
\cite{Polchinski:1987dy,Dorigoni:2009ra} in Sec.~\ref{GetReady}. We
will argue in Sec.~\ref{SECtraj} that, in theories with scale but not
conformal invariance, RG flows must have either limit cycles or ergodic
behavior.  We also make some general comments about the possibility of
uncovering fixed points with enhanced internal symmetries. We show (at least to second order in the loop expansion) that scale implies conformal invariance in two classes of models in Sec.~\ref{nogos}. This is interesting in its own right, but also sets
the stage for discovering models with scale but not conformal
invariance. An example of the latter we present in Sec.~\ref{TwoSOneF}. For conciseness we present explicitly the analysis and results for the simplest model
only, consisting of one Weyl spinor and two real scalars which also
displays a limit cycle with scaling symmetry. However, on this cycle
this model's scalar potential is unbounded from below. We close with a
short summary and a brief discussion of some interesting new open
questions.

\newsec{Preliminaries}[GetReady]
In order to establish notation and for completeness, we begin by
reviewing the conditions for scale and conformal
invariance\rcite{Callan:1970ze,Coleman:1970je,Polchinski:1987dy}. The
dilatation current is of the form
\eqn{
  \mathcal{D}^\mu(x)=x^\nu \T{\mu}{\nu}(x) -V^\mu(x)\,.}[DmuGiven]
Here $T^{\mu\nu}(x)$ is any symmetric stress-energy tensor and
$V^\mu(x)$ is any current that does not depend explicitly on
$x^\mu$. The freedom to choose among different symmetric stress
tensors is compensated through changes in the current $V^\mu$. The
improved stress-energy tensor can be particularly useful since it does
not get renormalized\rcite{Callan:1970ze}. Given a choice of stress-energy
tensor, scaling will be a symmetry if it is possible to find a current
$V^\mu$ such that
\eqn{
  \T{\mu}{\mu}=\partial_\mu V^\mu\,.}[DilInv]
Conformal invariance is equivalent to the existence of a traceless
stress-energy tensor. However, the stress-energy tensor in Eq.~\DilInv
need not be traceless. For $D>2$ it is sufficient that
\eqn{
  \T{\mu}{\mu}=\partial_\mu\partial_\nu L^{\mu\nu}}[ConfInv]
for some local tensor operator $L^{\mu\nu}$, for then one can
explicitly construct a traceless stress-energy tensor out of $T^{\mu\nu}$ and
$L^{\mu\nu}$. It follows that the condition that a model has scale but
not conformal invariance is to satisfy Eq.~\DilInv, with the
additional condition that the current $V^\mu$ cannot be written as a
conserved current $J^\mu$ plus the divergence of a two index symmetric
tensor $L^{\nu\mu}$,
\eqn{
  \T{\mu}{\mu}=\partial_\mu V^\mu,\text{ where }
  V^\mu\neq J^\mu+\partial_\nu L^{\nu\mu}\text{ with }\partial_\mu J^\mu=0.  }[DilNoConfCond]

Finding candidates for a current that one can use in the test
Eq.~\DilNoConfCond is not difficult. In $D$ space-time dimensions the
scaling dimension of the current must be $D-1$. In a perturbative setting with a collection of real
scalars,\foot{Indices from the beginning of the roman alphabet are in
  scalar-flavor space, while indices from the middle are in
  fermion-flavor space. In this work we don't consider gauge fields \rcite{Fortin:2011sz}.}  $\phi_a$, and Weyl spinors, $\psi_i$, the most general candidate is\rcite{Dorigoni:2009ra}
\eqn{
  V_\mu=Q_{ab}\phi_a\partial_\mu\phi_b-P_{ij}\bar{\psi}_i i\bar{\sigma}_\mu\psi_j.
}[CandidateV]
Without loss of generality one may take $Q_{ab}$ to be antisymmetric,
$Q_{ab}=-Q_{ba}$. Furthermore, in order for $V^\mu$ to be Hermitian,
$P_{ij}$ has to be anti-Hermitian, $P^\ast_{ij}=-P_{ji}$. The unknown
coefficients $Q_{ab}$ and $P_{ij}$ are to be determined by satisfying
Eq.~\DilNoConfCond. One may well expect that they depend on the
coupling constants of the model: the dimension of the operators
$\phi_a\partial_\mu\phi_b$ and $\bar{\psi}_i\bar{\sigma}_\mu\psi_j$
that go into $V_\mu$ is not generally $D-1$ in an interacting model,
so the coefficients may have to make up for the difference. But in a
perturbative model this difference is small. Hence, operators with
naive dimensions that differ from $D-1$ are not included in the
candidate current.

To proceed further we need to specify the model some more. In
order to have both UV and IR fixed points, so we may study the
relation between scale-invariant and conformal theories, we consider only
$D=4-\epsilon$ at small $\epsilon$. The action integral defines the
coupling constants as follows:
\eqn{
S=\int
\!d^Dx\left(\frac12\partial_\mu\phi_a\partial^\mu\phi_a
+i\bar{\psi}_i \bar{\sigma}^\mu\partial_\mu\psi_i
-\frac{\lambda_{abcd}}{4!}\phi_a\phi_b\phi_c\phi_d
-\frac{{y_{a|ij}^{\phantom{*}}}}{2}\phi_a\psi_i\psi_j
-\frac{{y_{a|ij}^*}}{2}\phi_a\bar\psi_i\bar\psi_j\right).}[Lag]

We can now test these models for scale and conformal invariance. The
trace of the energy momentum tensor is
\eqn{
\T{\mu}{\mu} = -\frac{\beta_{abcd}}{4!}\phi_a\phi_b\phi_c\phi_d
-\frac{{\beta_{a|ij}^{\phantom{*}}}}{2}\phi_a\psi_i\psi_j
-\frac{{\beta_{a|ij}^*}}{2}\phi_a\bar\psi_i\bar\psi_j,}
up to terms that vanish by the equations of motion.\foot{Equations of motion can be used and are not renormalized, see Ref.~\rcite{Politzer:1980me}.} The divergence of
the candidate current, after using the equations of motion, is
\eqn{
  \partial_\mu V^\mu=
  -\tfrac{1}{4!}\CQ_{abcd}\phi_a\phi_b\phi_c\phi_d
  -\tfrac12 (\CP_{a|ij}\phi_a\psi_i\psi_j+\text{h.c.}),}
where
\twoseqn{
  \CQ_{abcd}&=Q_{ae}\lambda_{ebcd}+\text{3 permutations},}[QDefd]
  {
    \CP_{a | ij}&=Q_{ab}y_{b | ij}+(P_{ki}y_{a | jk}+i\leftrightarrow j).}[PDefd][QPDefd]
With these results, condition \DilNoConfCond becomes
\twoseqn{
  \beta_{abcd}-\CQ_{abcd}&=0,}[ScInvQ]
{
  \beta_{a|ij}-\CP_{a|ij}&=0.}[ScInvP][ScInvII]
The problem has been reduced to solving these algebraic
equations. Here, by solving the equations we mean finding $Q_{ab}$ and
$P_{ij}$ that satisfy these equations for some values of the coupling
constants. To reiterate, the equations need not be satisfied
identically, that is, for every value of the couplings. Still the
equations are not trivial because there are more equations than free
variables in $Q_{ab}$ and $P_{ij}$, so generally we expect no
solutions or, if the theory has fixed points, a trivial solution in
which both $Q_{ab}$ and $P_{ij}$ vanish. If these are the only
solutions, then we have that scale implies conformal invariance.

\newsec{Trajectories and enhanced-symmetry fixed points}[SECtraj]

If a solution with $\beta\ne0$  is found, then there exist a
point in parameter space with scale but not conformal
invariance. Perhaps it is because such a point has never been found
that it is universally referred to in the literature as a fixed
point. But clearly the point cannot be fixed because the beta
functions do not all vanish there. The point must lie on an
RG trajectory. Physical properties are common to the complete
trajectory. It follows that the whole trajectory displays scale but
not conformal invariance.

RG trajectories are defined by
\eqn{
  \frac{d\bar \lambda_{abcd}(t)}{dt}=\beta_{abcd}(\bar\lambda_{abcd}(t),\bar y_{a|ij}(t))
  \qquad\text{and}\qquad
  \frac{d\bar y_{a|ij}(t)}{dt}=\beta_{a|ij}(\bar \lambda_{abcd}(t),\bar y_{a|ij}(t))\,,
}[runningGs]
but now Eqs.~\ScInvII hold along the whole trajectory. So, it must also be true that
\eqn{
  \frac{d\bar \lambda_{abcd}(t)}{dt}=\CQ_{abcd}(\bar\lambda_{abcd}(t),\bar y_{a|ij}(t))
\qquad\text{and}\qquad
  \frac{d\bar y_{a|ij}(t)}{dt}=\CP_{a|ij}(\bar \lambda_{abcd}(t),\bar y_{a|ij}(t))\,.
}[runningGbyQP]
Therefore, the running couplings solve both sets of equations \runningGs and \runningGbyQP simultaneously. This is a remarkable condition.

If the dependence of $Q_{ab}$ and $P_{ij}$ on coupling constants is
simple, one may integrate the equations readily. In fact, in the
explicit examples we give below, $Q_{ab}$ are constants (independent of
coupling constants) and $P_{ij}$ vanish. Hence, $\CQ$ and $\CP$ are
linear in the couplings. Moreover, since $Q_{ab}$ is real and
antisymmetric, it has purely imaginary eigenvalues, and thus the
trajectories must be periodic or quasi-periodic. These correspond
to the existence of limit cycles and ergodicity, respectively, in the
classification of possible behaviors of
RG trajectories\rcite{Wilson:1970ag,Wilson:1973jj,Lorenz:1963yb,TUS:TUS136}. There
do not seem to be any reported examples in relativistic field theory
displaying either of these. There appears to be a tight connection
between these behaviors and scale but not conformal symmetry, and no
examples of such models were known before this work\rcite{Fortin:2011sz}.

There is a simple reason to expect periodic or quasi-periodic
RG trajectories. The conserved dilatation current, Eq.~\DmuGiven, is a
combination of a scaling and a rotation in field space. The latter is
a transformation in a compact group. A curve in a compact space must
be periodic or quasi-periodic.  Now, a scale transformation gives a
translation along the RG trajectory.\foot{Except at a fixed point. We
  are considering the behavior of a model with scale but not conformal
  symmetry.} So as the RG trajectory is traversed, the field rotation
eventually goes back to the identity, or arbitrarily close to the
identity. Hence, the scale transformation itself must go back to, or
arbitrarily close to, the identity and therefore, by continuity in
parameter space, the RG trajectory must return to, or arbitrarily
close to, the starting point.

There may also be solutions to \ScInvII that have $\beta=0$ but for
which $Q\ne0$ or $P\ne0$. The trace of the stress-energy tensor
vanishes and from Eq.~\DilInv we see that $V^\mu$ is a conserved
current. This is the case of a fixed point with an enhanced internal
symmetry. On the trajectories that flow towards the fixed point there
is no symmetry associated with the current $V^\mu$, since away from the
fixed point $\T{\mu}{\mu}\ne0$.

\newsec{Scale implies conformal: two classes of models in
  \texorpdfstring{$4-\epsilon$}{4-epsilon}}[nogos]
We now prove, to second order in perturbation theory, that scale
implies conformal invariance for models of an arbitrary number of scalars in $D=4-\epsilon$. We also prove the same result \textit{to all orders in perturbation theory}
for models of an arbitrary number of Weyl spinors and exactly one real
scalar.

\subsec{Models of scalars}
Polchinski showed that, in a model of any number of scalar fields with
arbitrary quartic couplings, scale implies conformal invariance at
one loop. We review his argument and then extend it to second order in
perturbation theory.

The model is given by the Lagrangian in  Eq.~\Lag with all of the fermion
fields set to zero and $D=4-\epsilon$. Recall that we are trying to find a
solution to Eq.~\ScInvQ at a point in parameter space for which the
beta function does not vanish. Polchinski's argument is the
following. Using the explicit form of the $\beta$ function to one-loop
order,
\eqn{
  \beta^{(\text{1-loop})}_{abcd}=-\epsilon\lambda_{abcd}+
  \frac{1}{16\pi^2}\left(\lambda_{abef}\lambda_{efcd}+\text{2 permutations}
  \right)
}
and the explicit form of $\CQ$ given in Eq.~\QDefd, one can verify by
explicit computation that
\eqn{
 \CQ_{abcd}^{\phantom{a}}\beta^{\text{(1-loop)}}_{abcd} = 0\,.
}
From Eq.~\ScInvQ it follows that $\CQ_{abcd}\CQ_{abcd}=0$ which
implies that $Q_{ab}=0$, and thus leaves us only with fixed-point solutions to
Eq.~\ScInvQ.

We extend this argument to two loops by brute force. We have used the
two-loop expression for the beta function in dimensional regularization \rcite{Jack:1984vj} to verify that
$\CQ_{abcd}^{\phantom{a}}\beta^{\text{(2-loop)}}_{abcd} = 0$. We do not give the
details of the computation, there is little to be learned from the
explicit and lengthy expressions. One can verify, however, that with
all possible contractions of three $\lambda$'s with four free indices,
the only way to get $\CQ_{abcd}(\lambda\lambda\lambda)_{abcd}\neq 0$
is to contract two indices in the same $\lambda$. But the diagrams
that correspond to such a contraction are zero in dimensional
regularization, and so we find
$\CQ_{abcd}^{\phantom{a}} \beta^{\text{(2-loop)}}_{abcd}=0$. It follows that scale
implies conformal invariance at two loops too.

\subsec{Models of Weyl spinors and one real scalar} Dorigoni and
Rychkov have extended Polchinski's analysis to the case of models with
spinors and scalars described by the Lagrangian density in Eq.~\Lag,
with $D=4-\epsilon$\rcite{Dorigoni:2009ra}.  They showed that, at one
loop, for the conditions in Eqs.~\ScInvII to be satisfied one must set
both $\CP$ and $\CQ$ to zero. Hence, at any scale-invariant point one
must have vanishing beta functions.  They argue as follows.  First
contract Eq.~\ScInvP with $\CP^\ast_{a|ij}$ to obtain
\eqn{
  \CP^\ast_{a|ij}\beta^{\phantom{\ast}}_{a|ij}=\CP^\ast_{a|ij}\CP^{\phantom{\ast}}_{a|ij}.}[condPii]
The right-hand side of Eq.~\condPii is a real number. However, using the explicit form
of the beta function at one-loop order one finds that the real part of
the left-hand side of Eq.~\condPii is identically zero, and so $\CP_{a|ij}$
vanishes. Furthermore, contracting Eq.~\ScInvQ with $\CQ_{abcd}$ and using
$\CP_{a|ij}=0$ one finds that $\CQ_{abcd}$ vanishes as
well. Consequently, Eqs.~\ScInvII are satisfied only at conformal fixed points.

If we now attempt to extend this result by using the explicit form of
the beta function to two-loop order\cite{Jack:1984vj}, we find that
$\Rea (\CP^*\beta)$ does not vanish in the general case and, therefore, condition \condPii does not generally require the vanishing of
$\CP$. In the absence of a general argument we inspect specific cases.

Consider a model with an arbitrary number of spinors and only one real
scalar field. In this case the Lagrangian in Eq.~\Lag has only a single
scalar self-coupling $\lambda$, and the Yukawa couplings form a
single matrix $y_{ij}$. The $1\times1$ antisymmetric matrix $Q_{ab}$
vanishes and so do $\CQ$ and, by Eq.~\ScInvQ, $\beta_\lambda$. Hence
the condition for scale invariance is solved only if $\lambda$ is at a
fixed point. The beta function $\beta_{ij}$ for the Yukawa couplings
is also a single matrix and is given by the matrix product of $y$
times a real polynomial in $y^\dagger y$, with coefficients that are
real functions of $\lambda$. Therefore, to all orders in perturbation
theory, $\beta$ is of the form $Hy$ where $H$ is a Hermitian matrix,
$H^\dagger = H$, which satisfies $Hy=yH$. It follows that
$\CP_{ij}^\ast\beta_{ij}^{\phantom{\ast}}=\text{Tr}(\CP^\dagger\beta)=\text{Tr}[(P^\dagger
y^\dagger - y^\dagger P^T) H y]=-\text{Tr}[P (y^\dagger Hy + (yHy^\dagger)^T)] $, that is, the product of
the anti-Hermitian matrix $P$ with the Hermitian matrix
$y^\dagger Hy+(yHy^\dagger)^T$. Therefore, the trace is purely imaginary and we can now
complete the argument: the right-hand side of Eq.~\condPii, being a real
number, must vanish. Hence $\beta_y$ must vanish if the condition for
scale invariance, Eq.~\ScInvP, is satisfied. We have shown that, to all
orders, if there are scale-invariant points, they are also fixed
points.

\newsec{The simplest example}[TwoSOneF]

If we attempt to continue our analysis of explicit cases we find a
snag: we cannot complete the argument that $\CQ$ and $\CP$ vanish. In
a model with two (or more) real scalars and one (or more) Weyl spinors
the matrix $Q_{ab}$ does not automatically vanish, as was the case in
models with a single real scalar. Then, using the explicit form of the
two-loop beta function\cite{Jack:1984vj} for the model \Lag with at
least two real scalars and at least one Weyl spinor, one finds
that
 $
  \Rea(\CP^\ast_{a | ij}\beta^{\text{(2-loop)}}_{a | ij})\neq 0  $.
This does not mean that non-trivial solutions to Eq.~\ScInvII must
exist. But it indicates a direction  to investigate.

The simplest theory of this type has one Weyl spinor $\psi$ and two
real scalars $\phi_1$ and $\phi_2$. The Lagrangian is
\eqn{
  \mathscr{L}=\text{kin.\
    terms}-\frac{\lambda_1}{24}\phi_1^4-\frac{\lambda_2}{24}\phi_2^2-\frac{\lambda_3}{4}\phi_1^2\phi_2^2-\frac{\lambda_4}{6}\phi_1^3\phi_2
  -\frac{\lambda_5}{6}\phi_1\phi_2^3-\left(\frac{y_1}{2}\phi_1\psi^2+\frac{y_2}{2}\phi_2\psi^2+\text{h.c.}\right),
}[LagTwosOnef]
and the candidate for $V^\mu$ can be written as
\eqn{
  V_\mu=q(\phi_1\partial_\mu\phi_2-\phi_2\partial_\mu\phi_1)-p\bar{\psi}\bar{\sigma}_{\mu}\psi,
}[VmuTwosOnef]
where $q$ and $p$ are real numbers. The connection with the notation
for the general model of Eq.~\Lag is that
$\lambda_{1111}\equiv\lambda_1, \lambda_{1112}\equiv\lambda_4,
\lambda_{1122}\equiv\lambda_3, \lambda_{1222}\equiv\lambda_5$,
$\lambda_{2222}\equiv\lambda_2$ and $y_{a|11}\equiv y_a$.
Using the beta functions in
Ref.~\cite{Jack:1984vj} we find
\eqn{
\Rea(\CP^\ast_{a | ij}\beta^{\text{(2-loop)}}_{a | ij})=\frac{1}{(16\pi^2)^2}[
y_a y_b y^\ast_c y^\ast_d(Q_{ae}\lambda_{bcde}+Q_{ce}\lambda_{abde})-\tfrac{1}{24}y_a y^\ast_b(Q_{ac}\lambda_{bdef}+Q_{bc}\lambda_{adef})\lambda_{cdef}],
}[RePstarBeta]
which does not necessarily vanish. We proceed to search for
non-trivial solutions to \ScInvII.

In order to explain the strategy that we follow in solving \ScInvII it is
useful to sketch the form of these equations. Retaining up to two-loop contributions to the beta functions, we have, schematically,
\eqna{
  \beta_{abcd}-\CQ_{abcd}&\sim -\epsilon \lambda +
  \frac{\lambda^2+\lambda(y^*y)+(y^*y)^2}{16\pi^2}
  + \frac{\lambda^3+\lambda^2(y^*y)+\lambda(y^*y)^2+(y^*y)^3}{(16\pi^2)^2}
  -Q\lambda=0,\\
  \beta_{a|ij}-\CP_{a|ij}&\sim -\frac{\epsilon}{2} y +
  \frac{yy^*y}{16\pi^2}+ \frac{y(y^*y)^2+yy^*y\lambda+y\lambda^2}{(16\pi^2)^2}
  -Qy - Py=0.}[ScInvIIsketch]
The form of these equations suggests that we search for solutions as
an expansion in $\epsilon$. To lowest order the solutions should
correspond to the fixed points obtained from balancing the
``classical'' term, $\sim \epsilon\lambda$, against the first quantum
corrections, i.e.\ the one-loop terms. The Polchinski--Dorigoni--Rychkov
argument tells us that we should ignore $Q$ and $P$ at this order. So
we take
\begin{align}
\label{couplingEpsExp}
\lambda_{abcd} &=\sum_{n=1}^{\infty}
\lambda^{(n)}_{abcd}\epsilon^{n}, &
y_{a|ij} &=\sum_{n=1}^{\infty}
y_{a|ij}^{(n)}\epsilon^{n-\frac12},\\
\label{QPEpsExp}
Q_{ab} &=\sum_{n=2}^{\infty}
Q^{(n)}_{ab}\epsilon^{n}, &
P_{ij} &=\sum_{n=2}^{\infty}
P^{(n)}_{ij}\epsilon^{n}.
\end{align}
Notice that the nature of the expansions for the coupling constants
is dictated by the two lowest-order terms, that is, by the location
of a would-be fixed point. Meanwhile, the expansions for $Q$ and $P$
start at order $\epsilon^2$ or higher, since they must vanish if only
up to one-loop terms are retained in the beta functions.

For this particular model there are nine equations to solve,
corresponding to the beta functions for two complex $y$'s and five
real $\lambda$'s in Eqs.~\ScInvIIsketch. Adding $p$ and $q$ to the
list of coupling constants, there are eleven variables. Thus, the
system is under-constrained. This is as it should be if we are to have
solutions along trajectories, but it is not computationally
convenient. Instead it is best to fix some variables. We can set
$\Ima(y_2)=0$ by redefining the fields by a phase rotation of the Weyl
spinor.

There are many fixed-point solutions, and one must be careful to check
that a solution to \ScInvII is not also a fixed point, even for
nonzero $q$ and/or $p$. One must also check that the solution does not
give a scalar potential that is unbounded from below. Unfortunately, in the simplest
example with one Weyl spinor and two real scalars, we have not found
a scale-invariant point with a bounded-from-below scalar potential
that is not conformally invariant.  We have verified that such
scale-invariant points exist in more general models, e.g.,\ with two
Weyl spinors and two real scalars \rcite{Fortin:2011sz}, but, to keep the presentation
simple, we only display here the scale-invariant point (and the
trajectory it lies on) in the model with one Weyl spinor and two real
scalars. In an $\epsilon$-expansion we find the
scale-invariant solution
\eqn{\begin{aligned}
\lambda_1 &=\tfrac{821326-5427\sqrt{419802}}{607836}\pi^2\epsilon+\tfrac{518735723529516-118790842537\sqrt{419802}}{195971150186496}\pi^2\epsilon^2+\ldots,\\
\lambda_2 &=\tfrac{7(373922-141\sqrt{419802})}{607836}\pi^2\epsilon-\tfrac{23(6387330973\sqrt{419802}-5101825968812)}{65323716728832}\pi^2\epsilon^2+\ldots,\\
\lambda_3 &=\tfrac{469(222+\sqrt{419802})}{607836}\pi^2\epsilon+\tfrac{74835485902788+225616637735\sqrt{419802}}{195971150186496}\pi^2\epsilon^2+\ldots,\\
\lambda_4 &=\tfrac{7\sqrt{\tfrac{469}{74}(3601+6\sqrt{419802})}}{8214}\pi^2\epsilon\\
&\hspace{2cm}+\tfrac{\sqrt{\tfrac{7(2595761325955388328540064229507+4050673053526086225418178112\sqrt{419802})}{543700157374}}}{17272267776}\pi^2\epsilon^2+\ldots,\\
\lambda_5 &=\tfrac{67\sqrt{\tfrac{469}{74}(3601+6\sqrt{419802})}}{8214}\pi^2\epsilon\\
&\hspace{2cm}+\tfrac{\sqrt{\tfrac{67(668989476956566997057214743017+1051445250906514790976552640\sqrt{419802})}{170413482162}}}{17272267776}\pi^2\epsilon^2+\ldots,\\
y_1 &=\sqrt{2}\pi\sqrt{\epsilon}+\tfrac{1737927\sqrt{2}-3350\sqrt{209901}}{12616704}\pi\epsilon^{3/2}\\
&\hspace{2cm}+\tfrac{\sqrt{\tfrac{258756594352544587227002131-322169380386272743890676\sqrt{419802}}{3782}}}{1434065043456}\pi\epsilon^{5/2}+\ldots,\\
y_2 &=0,\\
q &=\tfrac{511\sqrt{\tfrac{469}{74}(3601+6\sqrt{419802})}}{33644544}\epsilon^3+\ldots,\\
p &=0.
\end{aligned}
}[ScaleInvPoint]

The reader will notice that the expansion for $q$ begins at third order in epsilon---this corresponds to the third order in the loop expansion of the beta function in Eqs.\ \ScInvII, which we have not included. It would seem that the solution is inconsistent and may disappear when the next order in beta is included. However, we have verified that this does not happen. That $q$ here begins at third order is due to a numerical accident of the two-loop Yukawa beta-functions.  All these issues will be discussed elsewhere.

\begin{figure}[ht]
\centering
\pstool{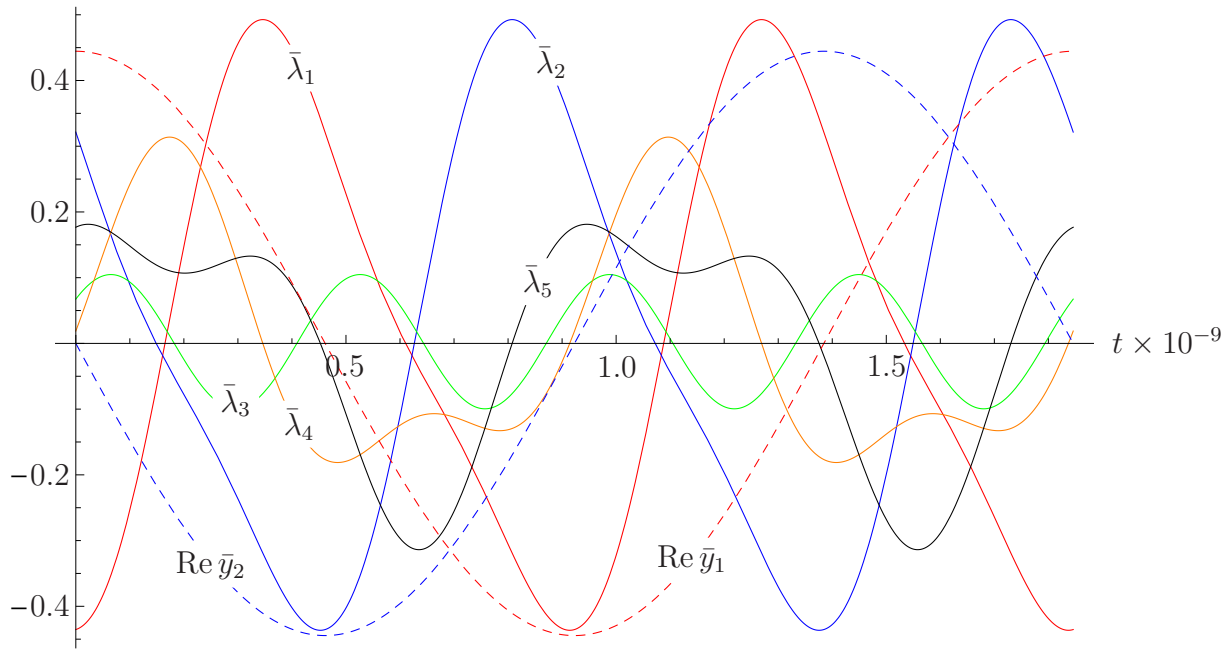}{\psfrag{t}{$t\times 10^{-9}$}
\psfrag{0.5}[][]{0.5}
\psfrag{1.0}[][]{1.0}
\psfrag{1.5}[][]{1.5}
\psfrag{-0.2}[][]{$-0.2$}
\psfrag{-0.4}[][]{$-0.4$}
\psfrag{0.2}[][]{0.2}
\psfrag{0.4}[][]{0.4}
\psfrag{l}[][]{\colorbox{white}{$\bar{\lambda}_1$}}
\psfrag{m}[][]{\colorbox{white}{$\bar{\lambda}_2$}}
\psfrag{n}[][]{\colorbox{white}{$\bar{\lambda}_3$}}
\psfrag{o}[][]{\colorbox{white}{$\bar{\lambda}_4$}}
\psfrag{p}[][]{\colorbox{white}{$\bar{\lambda}_5$}}
\psfrag{q}[][]{\colorbox{white}{$\Rea \bar{y}_1$}}
\psfrag{r}[][]{\colorbox{white}{$\Rea \bar{y}_2$}}
}

\caption{The couplings of the model with one Weyl spinor and two real scalars on an RG
  cycle, as a function of RG time. Here $\epsilon=0.01$ and the
  starting conditions are the solutions to \ScInvII given in the
  text, Eqs.~\ScaleInvPoint. }\label{coupplot}
\end{figure}

\begin{figure}[ht]
\centering
\includegraphics{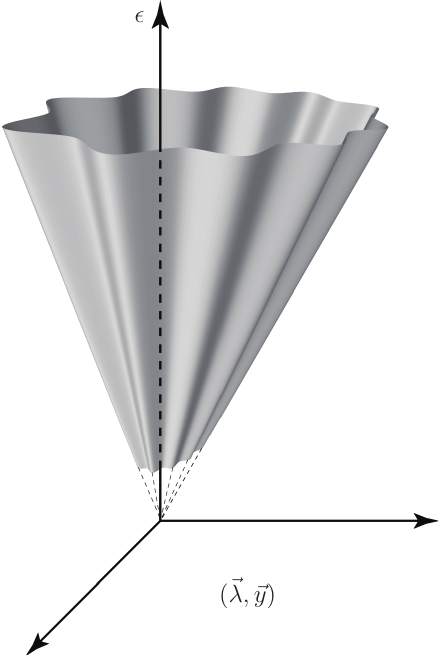}
\caption{Artistic rendition of scale-invariant trajectories as a function of $\epsilon$.}\label{hypercone}
\end{figure}

 Now that we have discovered a solution that is not a fixed point we
 can uncover the scale-invariant trajectory from
 Eqs.\ \runningGbyQP. For the scalar couplings, for example, organized as a
 five-dimensional vector, we have a system of coupled linear differential equations,
\eqn{
\frac{d\vec{\lambda}}{dt}={\SQ}\vec{\lambda},
}
where the matrix $\SQ$ is
\eqn{
\SQ=q\begin{pmatrix}
0 & 0 & 0 & 4 & 0\\
0 & 0 & 0 & 0 & -4\\
0 & 0 & 0 & -2 & 2\\
-1 & 0 & 3 & 0 & 0\\
0 & 1 & -3 & 0 & 0\\
\end{pmatrix}.
}
Once the matrix $\SQ$ is diagonalized the system decouples and is easy to solve. The solution, including the Yukawa couplings, is
\eqna{
\bar{\lambda}_1(t)&=\lambda_1 \cos^4 qt+\lambda_2\sin^4 qt+\tfrac{3}{2} \lambda_3 \sin^2 2qt + 4 \lambda_4\sin qt \cos^3 q t+4 \lambda_5 \sin^3 qt \cos qt,\\
\bar{\lambda}_2(t)&=\lambda_1 \sin ^4 q t+\lambda_2 \cos^4 q t+\tfrac{3}{2} \lambda_3 \sin^2 2 q t-4 \lambda_4 \sin ^3 q t \cos q t-4 \lambda_5 \sin q t \cos ^3 qt,\\
\bar{\lambda}_3(t)&=\tfrac{1}{4} \lambda_1 \sin ^2 2 q t+\tfrac{1}{4} \lambda_2 \sin ^2 2 qt+\tfrac{1}{4} \lambda_3 (3 \cos 4 q t+1)-\tfrac{1}{2} \lambda_4 \sin 4 qt+\tfrac{1}{2} \lambda_5 \sin 4 q t,\\
\bar{\lambda}_4(t)&=-\lambda_1 \sin q t \cos ^3q t+\lambda_2 \sin^3 q t \cos q t+\tfrac{3}{4}\lambda_3 \sin 4 q t+\tfrac{1}{2} \lambda_4 (\cos 2 q t+\cos 4 q t)\\
&\hspace{10cm}+\lambda _5 \sin^2 q t \,(2 \cos 2 q t+1),\\
\bar{\lambda}_5(t)&=-\lambda_1 \sin ^3 q t \cos q t+\lambda_2 \sin q t \cos ^3q t-\tfrac{3}{4}\lambda_3 \sin 4 q t+\lambda_4 \sin^2 q t \, (2 \cos 2 q t+1)\\
&\hspace{10cm}+\tfrac{1}{2}\lambda_5 (\cos 2 q t+\cos 4 q t),\\
\bar{y}_1(t)&=y_1 \cos q t+y_2 \sin q t,\\
\bar{y}_2(t)&=-y_1 \sin q t+y_2 \cos q t.
}[tDepSols]
Here the initial values for the scalar and Yukawa couplings on the right-hand side,
as well as the frequency $q$ are the solutions given in
Eqs.~\ScaleInvPoint above. As discussed above, the scalar potential
is unbounded from below for these values. Note that the imaginary
parts of the Yukawas vanish, hence the theory does not violate
CP. Note, furthermore, that these statements remain true throughout
the cycle, as indeed they should.  The couplings are plotted in
Fig.~\ref{coupplot}.
How do these solutions depend on $\epsilon$?
From Eqs.~\ScaleInvPoint it follows that the scale-invariant
trajectory disappears in $D=4$.  A sketch of the shrinking
trajectories in parameter space is shown in Fig.~\ref{hypercone}.

The $t$-dependent solutions \tDepSols can be obtained by replacing
\eqn{
\begin{pmatrix}\phi_1\\ \phi_2\end{pmatrix}
\to
\begin{pmatrix}\cos qt&\sin qt\\-\sin qt&\cos qt\end{pmatrix}
\begin{pmatrix}\phi_1\\ \phi_2\end{pmatrix}
}
in the Lagrangian \LagTwosOnef and then reading off the new coupling
constants as coefficients of the separate monomials of the potential
in $\mathscr{L}$. This is not surprising since the transformation
corresponds to the action of the charge $\int \! d^3x\, V^0$, and the
current in \VmuTwosOnef generates rotations among the two real
scalars.

We have also studied two additional cases. For a model of one Weyl spinor and three real scalars we find closed trajectories, but this model also has an unbounded scalar potential. For a model of two Weyl spinors and two real scalars we find scale-invariant trajectories with non-vanishing $Q$ and undetermined $P$. The potential is bounded from below and the model displays a limit cycle. The same limit cycle is found for a model consisting of two real scalars and one Dirac spinor (which dispels any concerns that may arise from the use of Weyl spinors in non-integral dimensions).  We will give the details of this model in a forthcoming publication \cite{Fortin:2011sz}.

\newsec{Discussion and conclusion}[Conclusion]
The dearth of examples of scale but not conformally invariant theories
has led to the general belief that scale implies conformal
invariance. As a result, a vast amount of knowledge has been amassed on
the behavior of both conformal theories and scale non-invariant
ones. We know virtually nothing about scale but not conformally invariant
theories. The examples we have found open the door for the exploration
of this radically new class of relativistic quantum field theories.

Are the trajectories we have found attractors, much like infrared
fixed points? What is the scalars' effective potential? Are there
models in physical space-time dimensions, i.e., integral $D$, which
exhibit limit cycles? What about supersymmetric models? Can one
systematize the search for models in $D=4-\epsilon$? What restrictions are imposed by scale invariance on Green's functions?  Are there phenomenological, model-building applications?  We plan to
address some of these questions in a forthcoming
publication\rcite{Fortin:2011sz}.

The question of models in integral dimensions is particularly
exciting. At the moment the best hopes are to either convincingly
extend our results here to $D=3$, much like in the theory of critical
phenomena, or to study $D=4$ models with a Yang--Mills fixed-point
coupling playing the role of $\epsilon$ in the $4-\epsilon$
examples. In light of our present results, we are happy to abandon the
gloomy pessimism of the past and plan to search for one such example in
earnest.

\ack{We thank Ken Intriligator and Julius Kuti for useful
  discussions. For tensor manipulations we have used
  \emph{Mathematica} with the package
  \href{http://www.xact.es/}{\texttt{xAct}}. This work was supported
  in part by the US Department of Energy under contract
  DOE-FG03-97ER40546.}

\bibliography{SvC_ref}

\end{document}